\documentclass[conference]{IEEEtran}
\IEEEoverridecommandlockouts
\usepackage{cite}
\usepackage{amsmath,amssymb,amsfonts}
\usepackage{algorithmic}
\usepackage{graphicx}
\usepackage{textcomp}
\usepackage{xcolor}
\usepackage{url}
\def\BibTeX{{\rm B\kern-.05em{\sc i\kern-.025em b}\kern-.08em
    T\kern-.1667em\lower.7ex\hbox{E}\kern-.125emX}}
\begin{document}

\title{A Survey on Recent Deep Learning-driven Singing Voice Synthesis Systems}

\author{\IEEEauthorblockN{Yin-Ping Cho, Fu-Rong Yang, Yung-Chuan Chang, Ching-Ting Cheng, Xiao-Han Wang, Yi-Wen Liu}
\IEEEauthorblockA{\textit{Department of Electrical Engineering} \\
\textit{National Tsing Hua University}\\
Hsinchu, Taiwan \\
yinping.cho@outlook.com; ywliu@ee.nthu.edu.tw}}

\maketitle

\begin{abstract}
Singing voice synthesis (SVS) is a task that aims to generate audio signals according to musical scores and lyrics. With its multifaceted nature concerning music and language, producing singing voices indistinguishable from that of human singers has always remained an unfulfilled pursuit. Nonetheless, the advancements of deep learning techniques have brought about a substantial leap in the quality and naturalness of synthesized singing voice.
This paper aims to review some of the state-of-the-art deep learning-driven SVS systems. We intend to summarize their deployed model architectures and identify the strengths and limitations for each of the introduced systems. Thereby, we picture the recent advancement trajectory of this field and conclude the challenges left to be resolved both in commercial applications and academic research.
\end{abstract}


\section{Introduction}
A singing voice synthesis (SVS) system is able to generate singing voice from a given musical score. For the future of music composition, one can imagine that a song can be listened to immediately 
after the song has been composed without recording. Recently, several approaches have been proposed to build a natural singing voice synthesis system in Japanese, English, Korean, Spanish, and so on \cite{sinsy, lstm, en}. Different from speech synthesis, the synthesized singing voice needs to follow the musical scores; performance of pitch and rhythm synthesis would directly influence the perceived quality. 

Before neural networks were widely used, unit concatenation \cite{cat} and hidden Markov Model (HMM) \cite{hmm} approaches were adopted for SVS. The unit concatenation synthesizer generates the singing voice by selecting the voice elements in the database and concatenation is performed consecutively. Commercially available tools such as Vocaloid \cite{vocaloid} and Synthesizer V\footnote{\url{https://synthesizerv.com/en/}} have successfully gathered loyal groups of users. In contrast, HMM-based SVS \cite{hmm} can model the spectral envelopes, excitation, and the singing voice duration separately. Then, speech parameter generation algorithms \cite{spg} are used to produce singing voice parameter trajectories. As HMM predates the advances in deep learning, the naturalness of HMM-based SVS is outperformed by what could now be achieved by neural networks.

Over the past few years, several types of neural networks have been employed for SVS, such as generic deep neural networks (DNN) \cite{dnn}, convolutional neural networks \cite{cnn}, a recurrent neural network with long-short term memory (LSTM) \cite{lstm}, and generative adversarial networks (GAN) \cite{gan}. Besides, taking advantage of the similarity to text-to-speech (TTS), some autoregressive sequence-to-sequence(Seq2Seq) models have been proposed \cite{byte, adv_kor}. In recent time, state-of-the-art deep learning architectures are adopted to tackle with the SVS task, such as the Transformer-based \cite{transformer} XiaoicSing \cite{xiao}, HifiSinger \cite{hifi} and diffusion denoising probabilistic model \cite{ddp} like DiffSinger \cite{diff}. However, these models typically need a large corpus for training; meanwhile, systems designed for lower data consumption, such as LiteSing \cite{lite} and Sinsy \cite{sinsy}, are now a heated research direction.

\begin{figure}[t]
    \centering
    \includegraphics[width=7.5cm]{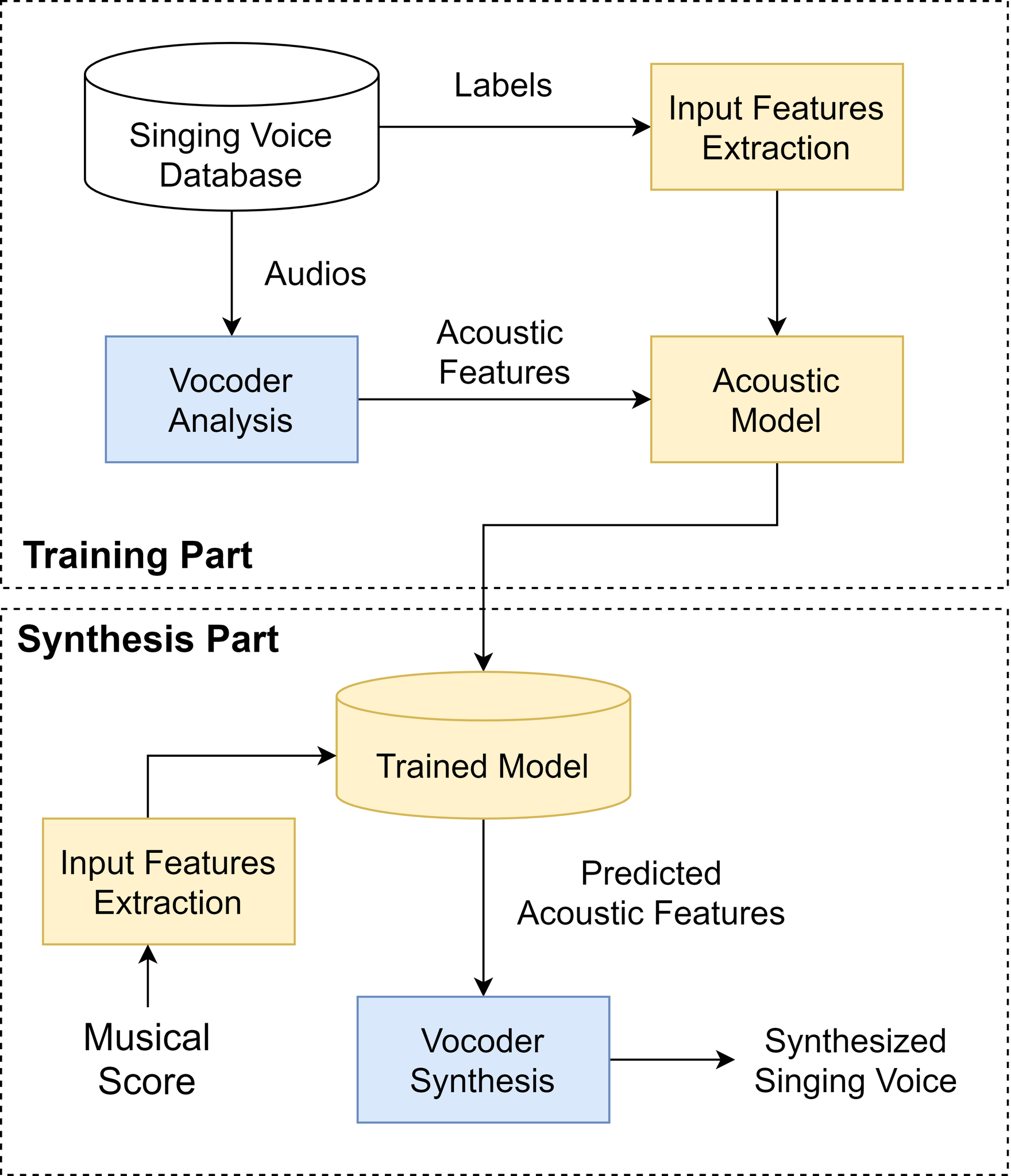}
    \caption{General architecture of recent deep learning-driven SVS systems.}
    \label{fig:arch}
    \vspace*{-3pt}
    \end{figure}

To mitigate the one-to-many difficulty of directly predicting the raw waveform from a musical score, recent deep learning-driven SVS systems mostly employ an acoustic model-vocoder architecture as in Fig~\ref{fig:arch}. In this setup, the vocoder maps frame-level parameters to the waveform, and the acoustic model only has to predict a parameter sequence with a length of the framed target waveform. This way, the mapping from the score to waveform is broken down into two simpler sub-tasks of lower dimension discrepancy between the mapping. Depending on the chosen vocoder, the frame-level synthesis parameters can be Mel-spectrograms \cite{hifi,diff} or parameters designed with more insights into the human voice, such as those with explicit and separate F0 values \cite{xiao,sinsy,lite}.

The rest of this paper is organized as follows: Section \uppercase\expandafter{\romannumeral 2} overviews three landmark deep learning SVS systems with high fidelity and naturalness. Two SVS systems with particular designs for low data resource training are described in section \uppercase\expandafter{\romannumeral 3}. Challenges and future directions of this topic are reported in Section \uppercase\expandafter{\romannumeral 4}, and the conclusions are summarized in \uppercase\expandafter{\romannumeral 5}.


\section{Naturalness and Audio Quality Milestones}
\subsection{XiaoiceSing: Transformer + WORLD}
XiaoiceSing \cite{xiao} was among the first of the deep learning-driven SVS systems that saw commercial deployment. It aimed to generate a singing voice with accurate pitch and rhythm that sounded natural and human-like. This system adopts the architecture of the FastSpeech \cite{fast} augmented with singing-specific modifications for the acoustic model and uses WORLD as the vocoder \cite{world}. The singing-specific modifications mainly concern the addition of note duration and note pitch information. To ensure the correctness of rhythm, the authors proposed adding a syllable-level duration loss instead of relying solely on the phoneme-level duration as the original FastSpeech. As for a more robust pitch, the note pitch residually connects to the F0 output; therefore, the acoustic model only has to predict the relative F0 variations in the human singing voice and add it onto the given note pitch contour. The experiments showed that XiaoiceSing outperformed the baseline system \cite{conv} in both subjective and objective evaluations. In particular, the A/B tests showed a dominant preference favoring XiaoiceSing in terms of F0 and rhythm naturalness, proving the effectiveness of the proposed modifications. However, the authors remarked that the mean opinions scores (MOS) of XiaoiceSing were held back by the waveform quality bound of the WORLD vocoder, although the acoustic model performed well in objective metrics. Also, this system consumed a large total of 74 hours of single professional singer’s training data, which would be inaccessible in a purely academic context.

\subsection{HiFiSinger: Transformer + Neural Vocoder}
Building on the foundation of XiaoiceSing, HiFiSinger \cite{hifi} aims to defy its waveform quality limitations. While HiFiSinger adopted the same FastSpeech-based acoustic model of XiaoiceSing, it swapped out WORLD and employed a Parallel WaveGAN (PW-GAN) \cite{pwg} neural vocoder to generate waveforms at a high-fidelity 48kHz sample rate. Therefore, the acoustic model is modified to predict a Mel-spectrogram as the main synthesis parameter and F0 contour with voiced/unvoiced (U/UV) flags as auxiliaries for the neural vocoder. To prompt better Mel-spectrogram fidelity, the authors proposed a sub-frequency GAN (SF-GAN) scheme where three discriminator networks operate on three overlapping sub-bands. This way, the GAN network should avoid the time-frequency resolution tradeoff. On the vocoder side, the main addition is the multi-length GAN (ML-GAN) that deploys multiple discriminator networks working on different waveform input lengths. This should make the discriminator handle both the long-time structures and short-time details better than the original single network. Also, predicted F0 and U/UV flags are integrated as auxiliary features for the modified PW-GAN alongside the Mel-spectrogram. The comparative mean opinion score ablation test confirmed that these modifications all positively contributed to the final system. In terms of perceived naturalness and audio quality, this system proved to be a significant improvement over XiaoiceSing in the MOS test. In the 48kHz generation configuration, HiFiGAN’s MOS (3.76) approached that of the ground truth high-fidelity recordings (4.03).

\subsection{DiffSinger: Denoising Diffusion Probabilistic Model + Neural Vocoder}
To further enhance the acoustic model’s prediction accuracy and robustness of Mel-spectrograms, DiffSinger \cite{diff} utilizes 
a generative model paradigm novel at the time of this writing: the denoising diffusion probabilistic model \cite{ddp}. Instead of directly optimizing the acoustic model to generate Mel-spectrograms, DiffSinger formulates the generation task into a parameterized Markov chain conditioned on the musical score. The diffusion process of this Markov chain gradually scales the Mel-spectrogram and applies noise until it becomes Gaussian noise. Conversely, the denoising process iteratively subtracts a portion of noise from the noisy input and rescales it until it becomes a Mel-spectrogram. To improve the speed and robustness of the denoising process, the authors proposed a shallow diffusion mechanism for denoising inference. This mechanism utilizes a simple auxiliary decoder seen in Transformer-based acoustic models \cite{xiao, hifi} to generate a rough Mel-spectrogram from the musical score. The denoising process can use this rough approximation as a starting point much closer to the Mel-spectrogram end of the Markov chain. Therefore, the inference process starts denoising from a noisy input with resemblance to the target Mel-spectrogram and requires much fewer steps to complete. The resulting acoustic model showed substantial quality improvements in MOS over its state-of-the-art counterparts in both text-to-speech (TTS) and SVS tasks. Also, with the shallow diffusion mechanism, it achieved a real-time factor (RTF) of 0.191 on a single RTX V100, meaning that it has real-time applicability.

\section{Towards Data Efficiency}

\subsection{Sinsy: DNN + Neural Vocoder}
Sinsy \cite{sinsy} is designed to synthesize singing voices at appropriate timing from a musical score. It contains four modules: 1) a ``time-lag model’’ which controls vocal start timing of each note for adapting human singing habits; 2) a ``duration model’’ which estimates phoneme durations for pre-expanding features into the frame level; 3) a DNN-based ``acoustic model’’ which plays the role of a mapping function from the score feature sequences to the acoustic feature sequences; 4) a PeriodNet ``neural vocoder’’ which generates time-domain waveform samples conditioned on acoustic features. Besides, Sinsy applied singing-specific techniques including accurately modeling pitch by predicting the residual connection between the note pitch and the output F0 in log scale, and the vibrato modeling which expresses fluctuations with the difference between the original F0 sequence and the smoothed one. Moreover, Sinsy proposed two automatic pitch correction strategies, the prior distribution of pitch and the pseudo-note pitch, to prevent singing voices from becoming out of tune. Sinsy adopted 1-hour Japanese children’s songs performed by a female singer for training. The mean opinion scores in subjective evaluation tests concluded that the proposed system could synthesize a singing voice with better start timing of vocal, more natural vibrato, and more accurate singing pitch.

\subsection{LiteSing: WaveNet + WORLD}
LiteSing \cite{lite} was designed to be a fast, high-quality SVS system with an efficient architecture that requires little training data. Instead of pursuing marginal audio quality gain with neural vocoders, LiteSing goes back to using WORLD as the vocoder. This is a conscious choice meant to utilize WORLD’s characteristic of separating instantaneous spectral envelopes from F0, making the prediction task simpler for the acoustic model. Furthermore, LiteSing employed a condition predictor that separately predicts dynamic acoustic energy, V/UV flags, and dynamic pitch curve, which means the decoder only has to predict the spectral envelope and aperiodic components with a much lower variance. That is, LiteSing disentangles the compound and complicated prediction task into well-defined and simpler sub-tasks for the acoustic model. Therefore, the authors could employ a relatively small and fast non-autoregressive WaveNet \cite{wavenet} as the model’s backbone and still expect robust and high-quality synthesis. In addition, the authors added a Wasserstein GAN (WGAN) \cite{wgan} for the predicted acoustic features to combat the over-smoothing problem typical in WORLD parameter prediction tasks. The experiments limited the dataset to 48 minutes of audio to test LiteSing’s data efficiency. In the MOS evaluation, LiteSing achieved a score (3.60) comparable to that of the WORLD-resynthesized human singer’s audio (3.86). More importantly, although the comparing FastSpeech2 baseline yielded a marginally higher score (3.63), LiteSing used only one-fifteenth of the number of parameters (3.8M vs. 57.0M), proving its superior data efficiency; also, LiteSing was significantly preferred over FastSpeech2 in the A/B test for expressiveness. These results demonstrated the quality and efficiency of the proposed acoustic model. However, the tradeoff of using WORLD was obvious: the original audio received a MOS of 4.20, substantially higher than the WORLD-resynthesized version (3.86), and the audio fully synthesized by LiteSing (3.60).

\section{Challenges}
\subsection{Data Efficiency}
In Table~\ref{tab:time}, we can observe a tradeoff between the amount of data and the synthesis quality. Systems that achieved nearly human-level quality \cite{xiao, hifi, diff} were those designed to be more data-driven and consumed multiple hours of a singing voice to train. On the other hand, although some systems may function with around or less than one hour of data \cite{sinsy, lite}, their measures of simplification usually come with degradations that sum to a compromised overall synthesis quality. Since composing a singing voice dataset requires high-fidelity recording equipment, a good singer, and substantial post-editing and score annotating efforts, data for SVS systems are costly to obtain. Therefore, the data efficiency of an SVS system may dictate its real-world applicability. To enhance data efficiency, the strategy usually combines two techniques: simplifying the neural network modules \cite{sinsy, lite} and decomposing the SVS task into lower-dimension sub-tasks by expert knowledge in singing voices \cite{sinsy}.

\begin{table}[h]
    \caption{Summary of data usage}
    \label{tab:time}
        \centering
        \begin{tabular}{c|c }
        \hline
        System & 
        Amount of Singing voice data consumed
        \\
        \hline
        \textbf{XiaoiceSing} & 74 hours \\
        \textbf{HiFiSinger} & 11 hours \\ 
        \textbf{DiffSinger} & 6 hours \\
        \textbf{Sinsy} & 1 hour \\ 
        \textbf{LiteSing} & 48 minutes \\
        \hline
        \end{tabular}
    \end{table}

\subsection{The lack of unified open datasets}
Producing an SVS dataset requires high cost of recording and annotating. Furthermore, due to copyright concerns, there are very few such datasets and those that are publicly available are usually confined to singing old folk songs without copyright \cite{csd,nit,vocalset}. Consequently, it is difficult for new researchers without resources to enter the field, and it is tricky for the community to objectively compare and evaluate the quality of different systems, as each of them is usually trained on a unique, proprietary dataset. Also, SVS systems that facilitate these open datasets may suffer from domain discrepancies, where the training dataset contains old folk songs while the songs of synthesis interest are of modern styles. As a comparison, text-to-speech (TTS) is a generation task with similar characteristics and complexity; however, an ample amount of free, open datasets is available for TTS \cite{openslr,ljspeech}, which allows the field to progress rapidly and see a wide array of successful architectures and readily available commercial products. Although this issue is not a technical one, it is a significant limiting factor for the present and the prospect of SVS research. 

\subsection{The lack of interpretability and transparency}
Deep learning has pushed the envelope of SVS systems in terms of synthesized audio quality. Nevertheless, deep learning systems’ complexity makes it almost impossible to analytically understand the learned mapping from the input musical score to the end output waveform. This characteristic prevents researchers from extracting knowledge about the mechanisms that comprise the process of singing. Consequently, it substantially weakens the motivation of doing SVS research as a reverse-engineering means looking to gain insights about human vocalization and perception of music. Although this issue does not prevent application-driven research, it terminates these works’ potential to extend into fields such as medicine or psychoacoustics. Considering the high cost of constructing SVS systems, improving the interpretability and transparency of deep learning-driven SVS systems is worthy of research effort.

\subsection{Absence of emotion and singing technique control and variations}
The high variation and diversity of emotion and vocalization techniques are quintessential to singing voices. However, although state-of-the-art deep learning-driven SVS systems can synthesize singing voices with reasonable naturalness and sound quality, they offer no means to condition these systems to synthesize with specific emotions or singing techniques. This is an advanced issue we surmise will become a research hot spot for two reasons: 1) In a commercial context, being able to sing according to the user’s desired style for the songs is basic functionality of professional singers. Without similar controllability or customizability, an SVS system lacks completeness to enter the real-world market. 2) This is an issue entangled with data efficiency. It is proven in a similar high-variance speech synthesis research \cite{gst} that conditioning the system on implicit features such as emotion and prosodic styles are instrumental to enhancing the system's robustness and naturalness given the same training data. In cases that these implicit features are statistically dominant to the synthesis target, the absence of these conditions may lead to failure of convergence. Therefore, we are confident that this issue is one that should be put in the spotlight of future SVS research.

\section{Conclusions}
In this work, we reviewed some of the deep learning-driven SVS systems that are representative of this research topic. We have shown that these methods have demonstrated synthesis quality and naturalness comparable to real human singers. Despite these achievements, challenges yet to be resolved for this topic were identified --- namely, the need for better data efficiency of the systems, the lack of open and unified benchmark datasets like those available in TTS research, deep neural networks’ inherent absence of interpretability, and the lack of control and explicit variations of emotion and singing techniques quintessential for singing voice synthesis. These issues hurdle the commercial deployment of SVS systems and are of high academic interest. Therefore, we can expect them to be the center of advancements in the coming times.





\end{document}